# Quantum Hall effect and Landau levels in the 3D topological insulator HgTe

J. Ziegler[1], D.A. Kozlov[2,3], N.N. Mikhailov[3], S. Dvoretsky[3], D. Weiss[1]

[1] Experimental and Applied Physics, University of Regensburg, D-93040 Regensburg, Germany
[2] A.V. Rzhanov Institute of Semiconductor Physics, Novosibirsk 630090, Russia
[3] Novosibirsk State University, Novosibirsk 630090, Russia

**Abstract**

We review low and high field magnetotransport in 80 nm-thick strained HgTe, a material that belongs to the class of strong three-dimensional topological insulators. Utilizing a top gate, the Fermi level can be tuned from the valence band via the Dirac surface states into the conduction band and allows studying Landau quantization in situations where different species of charge carriers contribute to magnetotransport. Landau fan charts, mapping the conductivity $\sigma_{xx}(V_g, B)$ in the whole magnetic field – gate voltage range, can be divided into six areas, depending on the state of the participating carrier species. Key findings are: (i) the interplay of bulk holes (spin-degenerate) and Dirac surface electrons (non-degenerate), coexisting for $E_F$ in the valence band, leads to a periodic switching between odd and even filling factors and thus odd and even quantized Hall voltage values. (ii) A similar though less pronounced behavior we found for coexisting Dirac surface and conduction band electrons. (iii) In the bulk gap, quantized Dirac electrons on the top-surface coexist at lower $B$ with non-quantized ones on the bottom side, giving rise to quantum Hall plateau values depending – for a given filling factor - on the magnetic field strength. In stronger $B$ fields, Landau level separation increases, charge transfer between different carrier species becomes energetically favorable and leads to the formation of a global (i.e. involving top and bottom surface) quantum Hall state. Simulations using the simplest possible theoretical approach are in line with the basic experimental findings, describing correctly the central features of the transitions from classical to quantum transport in the respective areas of our multicomponent charge carrier system.

**Introduction**

The quantum Hall effect (QHE) [1] and Shubnikov-de Haas (SdH) oscillations with zero resistance states [2] are hallmarks of two-dimensional electron (2DES) or hole (2DHS) hole systems, realized, e.g., in semiconductor heterostructures [3] or in graphene [4]. These phenomena are closely connected to the discrete Landau level (LL) spectrum of charge carriers in quantizing magnetic fields. These Landau levels are spaced by the cyclotron energy $\hbar\omega_c$ ($\hbar$ is the reduced Planck constant and $\omega_c$ the cyclotron frequency) and form, as a function of magnetic field and Landau level index $n$, the Landau fan chart. In conventional two-dimensional systems one type of charge carriers, i.e., electrons or holes prevail, giving rise to one Landau fan chart and a regular sequence of SdH-peaks or quantum Hall plateaus, which occur in equidistant steps on an inverse magnetic field scale, $1/B$. A more complicated situation arises when two-dimensional electron and hole systems coexist like in heterojunctions with broken gap (type III heterojunction), e.g., in InAs/GaSb quantum wells [5]. This gives rise to hybridization [6] and a complicated interplay between LLs [7]. With the advent of three-dimen-sional (3D) topological insulators (TI) [8–11] a new class of two-dimensional electron systems appeared on the scene where the two-dimensional charge carrier system consisting of Dirac fermions forms a closed surface "wrapped" around the insulating bulk. In contrast to a conventional two-dimensional electron gas, these Dirac surface states are non-spin degenerate and have their spin orientation locked to their momentum. The unusual topology of the two-dimensional Dirac system creates a manifold of possibilities how the charge carriers on top and bottom surface interact for different positions of Fermi level and magnetic field strength. For a Fermi-level position in the conduction band, e.g., three different charge carrier species with different densities and mobilities exist: bulk electrons, which are spin degenerate and non-degenerate Dirac surface electrons on top and bottom surface. Here, we ignore charge carriers on the side facets, parallel to the applied magnetic field, which play only a minor role in the context of the present investigations. While SdH oscillations and the QHE have been observed in various TI materials [12–17] strained HgTe, a strong topological insulator [18], is in so far special, as it features unprecedented high mobilities $\mu$ with $\mu B \gg 1$ at magnetic fields as low as 0.1 T. This material thus serves as a model system to explore Landau quantization and magnetotransport in a situation where different types of charge carriers exist together. However, the physics discussed below is also valid for other topological insulator materials in which different kinds of charge carriers coexist at the Fermi level.

The first observation of the QHE in strained films of HgTe was reported in Ref. [12], which also presents calculations of the band dispersion of the surface states on top and bottom surface. Corresponding ab-initio calculations computing the band gap in which the TI surface states reside as a function of strain (i.e. lattice mismatch between HgTe and substrate) were presented in Ref. [19]. Typical band gaps for HgTe on CdTe are of the order of 20 meV. Since in the initial experiment the position of the Fermi level $E_F$ was unknown, we explored the QHE effect and SdH oscillations systematically as a function of a top gate voltage [13]. The top gate enables tuning the Fermi-level from the valence band, through the gap region with gapless surface states into the conduction band. As the QHE is also observed when the Fermi level is supposedly in the conduction band, Brüne et al. claimed that due to screening effects the Fermi level is pinned in the band gap so that only Dirac surface states exist at $E_F$ [20]. However, the underlying "phenomenological effective potential for the sake of keeping the Fermi level within the bulk gap is not proper" [21]. Below we show not only that the Fermi level can be easily tuned from the valence band into the conduction band, but we provide a comprehensive picture of Landau quantization and quantum Hall effect in the multi-carrier system of a topological insulator.

**Samples' Characterization**

The 80 nm-thick HgTe material, investigated here, has been grown by molecular beam epitaxy on (013)-oriented (GaAs); similar wafers have been used previously to study magnetotransport [13], cyclotron resonance [22], or quantum capacitance [23]. A $\pi$-phase shift of quantum capacitance oscillations [23], geometric resonances in antidot arrays [24], and the subband spectrum of HgTe-nanowires made of this material [25], confirm the topological nature of the surface states. As shown in Fig. 1(a), the heterostructure includes a 20 nm-thick $Cd_xHg_{1-x}Te$ buffer layer on either side and a 40 nm-thick CdTe cap layer to protect the pristine HgTe surfaces. A top gate stack consisting of 30 nm $SiO_2$, 100 nm $Al_2O_3$ and Ti/Au enables control of the charge carrier density and thus of the Fermi level via gating. Thus, the Fermi energy $E_F$ can be tuned from the valence band (VB) into the bulk gap and further into the conduction band (CB) [13,21,23]. For magnetotransport measurement we use a standard Hall bar geometry of length 1100 µm and width 200 µm, sketched in Fig. 1(b), and temperatures of T = 50 mK; the magnetic field *B* points perpendicular to the sample plane. We use a low AC current of 10 nA flowing through the Hall bar to prevent heating of the carriers. To measure longitudinal and transversal resistivities $\rho_{xx}$ and $\rho_{xy}$, respectively, we use standard low-frequency lock-in techniques.

Figure 1(c) shows low-field SdH oscillations taken at different gate voltages, i.e. Fermi level positions. At $V_g$ = -1.5 V the Fermi level is in the valence band while for +1V it is in the bulk gap; a simplified band structure is shown in Fig. 2(a) for guidance. Fig. 1(d) displays the typical $V_g$-dependence of the resistivity $\rho_{xx}$, measured at $B = 0$, and of the Hall resistance $\rho_{xy}$, taken at 0.5 T. The traces are very similar to the ones we reported previously [13,23–26]. The $\rho_{xx}$-trace exhibits a maximum at $V_g = -0.15$ V near the charge neutrality point (CNP, at $V_g$ = 0.25 V), and two characteristic weak humps at 0.4 V and 1.25 V. The latter are due to enhanced scattering of surface electrons and bulk carriers at the band edges, i.e., valence band (VB, marked by $E_V$ on $V_g$ axis) and conduction band (CB, marked $E_C$) [13]. Near the CNP the Hall resistance $\rho_{xy}$ changes sign as the majority charge carrier type changes from electrons to holes when $E_F$ enters the VB. Closer examination of magnetotransport data using the Drude model and the periodicity of SdH oscillations, as in Refs. [13,23], reveals partial densities and precise values of $E_V$ and $E_C$. The result of such analysis is shown in Fig. 1(e). Above $E_V$, electron densities $n_s$ are extracted from the slope of the Hall resistance ($n_s^{Hall}$) and from the periodicity of the SdH oscillations. SdH oscillation periods reveal different carrier densities, depending on whether the data is taken at low ($n_s^{SdH,low}$) or high magnetic fields ($n_s^{SdH,high}$). While $n_s^{Hall}$ and $n_s^{SdH,high}$ represent the total charge carrier density of the system (and therefore coincide), $n_s^{SdH,low}$ represents the top surface electrons only [13,23]. The reason for the latter is that the SdH oscillation of the bottom side electrons are at these small magnetic fields not yet sufficiently resolved, in contrast to the ones on the top surface. We will address this topic further below. An example of low-field SdH oscillations, taken at different gate voltages, is shown in Fig. 1(c). The traces exhibit pronounced oscillations starting below 0.5 T, indicating the high quality of the HgTe layer. Below $E_V$, surface electrons and bulk holes coexist [see, e.g. Fig. 2(a)] and the corresponding densities can be extracted using a two-carrier Drude model [13]. Extrapolating $p_s^{Drude}(V_g)$ in Fig. 1(e) to $V_g$ = 0 yields the exact $V_g$-position of $E_V$, and coincides with the $\rho_{xx}$ hump, mentioned above. The change in slope of the top surface carrier density $n_s^{SdH,low}$ at $V_g$ = 0 signals that $E_F$ crosses the conduction band edge [13,23].

Figure 2(b) demonstrates that the system displays a well-developed quantum Hall effect. For negative gate voltages $V_g$, the quantized plateaus stem from bulk holes. For positive $V_g$, the slope of $\rho_{xy}$ changes sign and the plateaus stem from Dirac surface states with contributions from conduction band states at higher positive $V_g$. At positive or negative bias voltages the band bending is expected to confine the bulk states at the surface so that they form a conventional two-dimensional electron or hole gas. Thus, the notion bulk electrons or bulk holes refers to states which are, when biased, largely confined at the surface as is sketched in Fig. 4(a).

**Experimental Data Overview**

To get further insight into the interplay of the different carrier species in the system we take a closer look on the formation of Dirac state LLs and their interaction with bulk carriers' LLs. To this end, we measured the magnetotransport coefficients $\rho_{xx}$ and $\rho_{xy}$ at a temperature of 50 mK and magnetic fields up to 12 T. The measured resistivity values were converted into conductivities by tensor inversion. The corresponding data set is plotted in Fig. 3(a) as a normalized conductivity $\sigma_{xx}(V_g, B)$ where each horizontal cut was normalized to its $B$-field average. Lines of high conductivity are colored from yellow to light blue and represent Landau levels, while deep blue color corresponds to the gaps between LLs. The investigated $V_g$-range spans Fermi level values from inside the valence band to deep into the conduction band. In strong magnetic fields, the data shows both $n$- and $p$-type Landau fans, almost symmetrical with respect to the CNP, at which surface and bulk electrons counterbalance bulk holes. The well-resolved $\nu = 0$ state is either a magnetic field induced insulator state at the CNP or can be viewed as formation of counter-propagating edge states [27–29]. The nature of this state is out of scope of the present work.

To increase resolution and visibility in the low-field region, we present in Fig. 3(b) the second derivative $\partial^2 \sigma_{xx}/\partial V_g^2$ of the above data for smaller ranges $B < 5$ T and $V_g = $ -2…2 V. Here, yellow regions of negative curvature of $\partial^2 \sigma_{xx}/\partial V_g^2$ correspond to maxima in the conductivity, i.e. the position of LLs. Green and blue areas in between represent gaps. The respective data uncovers a complex oscillation pattern with splitting, kinks and crossings of LLs. These features occur in different sections of the $(V_g, B)$ parameter space, indicating that several groups of carriers are involved in the formation of LLs. The analysis of the data can be done in several ways. One option is to exactly calculate LL positions using the k·p method and trying to fit the whole set of data. However, such an approach entails significant computational difficulties because Schrödinger and Poisson equations need to be calculated self-consistently for every point in the $(V_g, B)$ parameter space. Still, this procedure would not promote true insight into the underlying physics. Instead, we introduce a minimal but adequate computational model, which semi-quantitatively agrees with the experimental data. While a detailed description of the model is presented in the Appendix, we address here only the crucial points: Within the model, each group of carriers is characterized by its own set of LLs, typified by the charge they carry (empty LLs bear no charge, while occupied ones are negatively or positively charged), Landau degeneracy (spin-degenerate or not), LL dispersion and LL broadening. Further, a joint Fermi level $E_F$ and the total charge carrier density $(n_s - p_s)$, which depends linearly on $V_g$, characterize the entire system. The total charge is distributed among all LLs, determined by the position of the Fermi level. Here, we need to take into account the electrostatics of the system: changes of the electric fields in the device due to changes of the carrier densities within the subsystem via the gate shifts the band edges and thus the corresponding origins of the LLs. The effect of electrostatics on the position of LLs for several gate voltages is shown in Fig. 4, described in more detail in the caption and in Fig. S1 in [30]. It is important to note that the band edges and thus the origin of the LL fan charts depend on the gate voltage $V_g$. The output of the calculation is the position of the Fermi level, both partial $\nu_i$ and total filling factors $\nu$, partial densities of states (DOS) $D^i$ and Hall conductivities $\sigma_{xy}^i$ for specified $B$ and $V_g$ values. Figure 3(c) shows an example of such a calculation which displays the total density of states as a function of $V_g$ and $B$, thus reconstructing the LLs within the $(V_g, B)$ map. Above 4 T on the hole side and above 6-8 T on the electron side the calculated LLs form simple fan charts which both emanate from the CNP. This is in excellent agreement with the experiment in Fig. 3(a).

The combined analysis of experimental data and simulations allows us to classify six distinct regions within the $\sigma_{xx}(V_g, B)$ map, shown in Fig. 3(b). We will discuss the regions marked from 1 to 6 below with the help of our model calculations. Region 1 covers the high magnetic field region where the system is fully quantized, so that all LLs are resolved. Here, the total density $n_s - p_s$ determines the filling factors and the Landau level positions. By extrapolating the LLs in Fig. 3(a) from high fields down to $B = 0$ (not shown) all the lines meet at the CNP. Conversely, region 6 located at fields below 0.4 T, describes fully classical and diffusive transport. At intermediate magnetic fields, i.e. in large parts of regions 2-5, only some of the LLs, e.g. on the top surface, are resolved so that combinations of quantized and diffusive transport arise. In region 4, $E_F$ is in the gap so that only topological states on top and bottom surfaces are present at the Fermi level, but no bulk states. In region 2 & 3, bulk holes dominate the Landau spectrum but coexist with surface electrons on top and bottom. The coexisting surface electron and bulk hole Landau fans cause an intricate checkerboard pattern in region 3. In region 5, the surface states coexist with bulk electrons that are filled when $E_F > E_C$. Below, we consider these regions in more detail.

**Weak & Strong Magnetic Field Limits**

The behavior of the system in the limits of small and large magnetic fields is easily accessible. At small magnetic fields, i.e. in region 6 of Fig. 3(b), the LL separation is smaller than the LL broadening and transport is fully diffusive. In this regime the classical multi-component Drude model characterizes transport. The model assigns each group of carriers, labelled by index $i$, its own set of density $n_{s,i}$ and mobility $\mu_i$. The resulting total conductivity is then the sum of the partial conductivities [13,31]. This model explains the observed effects, including the nonlinear $\sigma_{xy}(B)$ trace as well as a distinct positive magnetoresistance when bulk holes and surface electrons coexist in the valence band [13]. The very low density of bulk carriers in the gap is reflected by the darker blue color (i.e. low $\sigma_{xx}$) with a sharp transition at the conduction band edge $V_g(E_C)$ and a smoother one at $V_g(E_V)$ in Fig. 3(a). The color difference between gap and bulk bands regions reflects the effect of classical localization of charge carriers described by $\sigma_{xx,i} \propto n_{s,i}/\mu_i B^2$ in classically strong fields ($\mu B \gg 1$, here typically valid for $B > 0.1$ T). Thus, the magnetoconductivity $\sigma_{xx}$ is lowest in the gap where the bulk carrier concentration with low mobility is significantly lower than the one in conduction band. The contrast of the color at $E_C$ in Fig. 3(a) directly confirms that the Fermi-level can be tuned into the conduction band, in contrast to earlier claims [20].

The opposite case of strong magnetic fields is characterized by completely (spin) resolved LLs and well-defined QHE state. On the electron side, e.g., three fan charts - one for the top surface, one for the bottom surface and, for the Fermi-level in the conduction band, one for bulk electrons – coexist. For constant carrier density (gate voltage), the Fermi-level jumps with increasing magnetic field from a fully occupied LL of the top surface electrons to the next lower one, belonging e.g. to the electrons on the bottom surface. This process is connected with a transfer of charges from the top to the bottom surface. The SdH oscillations reflect, in this case, the total carrier density of bottom plus top surface electrons. The total charge density $n_s - p_s$ controls the total filling factor $\nu$ and the position of the Landau levels. On the electron side, this means that the total electron density (bulk plus surface density) determines the filling factor, while in the valence band (where surface electrons and bulk hole coexist) it is the difference of electron and hole densities. The Landau fan chart, constructed from high field $\sigma_{xx}$ data in Fig. 3(a), is symmetric with respect to the CNP and periodic in $V_g$ and $1/B$. The entire high field LLs in Fig. 3(a), extrapolated to zero magnetic field, have their origin at the CNP. The pattern is very similar to that observed in other electron-hole systems, e.g., in graphene [4,32], but without graphene's spin and valley degeneracy. However, there is a quantitative asymmetry between electrons (positive filling factors) and holes (negative filling factors): the conductivity minima for electrons are deeper than for holes; at the same time, the conductivity maxima are broader for holes. This asymmetry reflects the difference of effective masses and cyclotron gaps, which differ by one order of magnitude [22,33] between the carrier types.

**Coexistence of quantized and diffusive carriers in the bulk gap.**

Here we focus on the bulk gap region, labeled as region 4 in Fig. 3(b) and magnified in Fig. 5(a). In this region, the Fermi level is in the bulk gap and only Dirac surface states contribute to transport. The SdH oscillations appear from 0.3 T on and show a uniform pattern, strictly periodic in $V_g$ and $1/B$, until a magnetic field of $B = 1.5...2$ T is reached. This regular behavior reflects that only the motion of one carrier species, i.e. the electrons on the top surface, is quantized [13,23]. The electrons on the bottom surface have lower density and mobility and their LL spectrum is not yet resolved. The bottom electrons thus contribute a featureless background to the conductivity. The $\sigma_{xy}(V_g)$ data in Fig. 5(c) clearly show this diffusive background: The individual low field $\sigma_{xy}(V_g)$ traces show pronounced non-quantized Hall plateaus with values, which increase with increasing $B$-field [see red dashed lines in Fig. 5(c)]. This behavior can be reproduced by adding the Hall conductivities of top and bottom surface electrons: $\sigma_{xy}^{tot} = \sigma_{xy}^{quant} + \sigma_{xy}^{diff} = \nu^{top} e^2/h + \mu\, n_s^{bot}\, e\, \mu B/(1 + (\mu B)^2)$. Here, $\sigma_{xy}^{quant}$ is the quantized Hall conductivity of the top surface, depending on the corresponding filling factor $\nu^{top}$ and the carrier density of the back surface, $n_s^{bot}$, which depends only weakly on the gate voltage. We estimate the increase of the plateau values with increasing $B$ in [30], which is in good agreement with experiment.

The SdH oscillations occurring below 1.5 T are described by a single fan chart, shown as dark blue lines in Fig. 5(a). The fan chart emerges at $V_g = 0.05$ V, corresponding to a position of the Fermi-level in the valence band. This point can be viewed as virtual Dirac point, at which the density of the top surface Dirac seemingly becomes zero. However, this would only be the case if the filling rate of the top surface electrons were constant, which is not the case. When $E_F$ enters the valence band, the filling rate decreases precipitously and the fan chart no longer fits the observed $\sigma_{xy}$ maxima [see dark blue dashed lines in the Fig. 5(a)]. The actual gate voltage at which $n_s^{top} = 0$ holds is located much deeper in the valence band and labeled $E_{DF,top}$, shown in Fig. 5(c). Typical partial densities of the different charge carriers and the corresponding filling rates, extracted from our experiments, are sketched in Fig. S1(a) and (b).

Below a magnetic field strength of about 1.5 T, the surface electrons on the bottom side do not contribute to the oscillatory part of $\sigma_{xx}$. This is an important distinction to other well-studied two-component systems like wide GaAs QWs [34–36], where SdH oscillations of different frequencies appear on a $1/B$ scale. At larger fields ($B > 2$T), bottom surface electrons contribute to quantum transport. This results in extra structure in the SdH-oscillations, which breaks the regularity of the fan chart of Fig. 5(a). In this regime the $\sigma_{xy}$ plateaus, shown in Fig. 5(c), become quantized in units of $e^2/h$. The data suggest that the transition from classical to quantum transport appears at different magnetic fields for top and bottom surfaces (0.3 T and 1.5 T, respectively). This might be caused by significantly different levels of LL broadening for electrons on top and bottom surfaces, however, this is not obligatory. The simulation shown in Fig. 6(a) shows that the fan chart of the top surface dominates even if one assumes the same disorder and mobility for top and bottom surface electrons. Fig. 6(a) shows the total density of states of top surface and bottom surface states. The dark blue fan chart, however, is the one of the top surface only but matches the total DOS perfectly. The fan chart of the back surface has observably different slope (see Fig. S2(b) in [30]). However, the top surface dominates the overall density of states because of its 2.5 times higher carrier density and thus higher SdH oscillation frequency.

**The Valence Band: Spin degenerate holes and spin-resolved electrons**

The most striking feature in transport and thus in the LL fan chart arises from coexisting surface electrons and bulk holes in regions 2 and 3 [see Fig. 3(b) and magnification in Figs. 5(b) and 5(e)], each characterized by its own set of LLs. Note that in the valence band both top and bottom surface electrons are present. However, only the top surface electrons participate in the formation of the observed LLs. The bottom electrons have an extremely small filling rate (experimentally indistinguishable from zero) and therefore do not develop observable Landau fans. The electron Landau levels start from the Dirac point of the top surface electrons, marked by $E_{DF,top}$ on the gate voltage scale of Fig. 5(b), and fan out towards positive gate voltages. The hole Landau levels, on the other hand start at the valence band edge $E_V$ [see Fig. 2(a) and Fig. 5(b)] and fan out towards negative gate voltages. The simultaneous filling of the two sets of Landau levels results in the formation of a checkerboard like pattern where minima and maxima alternate, clearly seen in the $\partial^2 \sigma_{xx}/\partial V_g^2$ color map of Fig. 5(b). This checkerboard pattern is very similar to the one observed in InAs/GaSb based electron-hole systems [7]. The checkerboard pattern is connected to an anomaly in the $\sigma_{xy}$ traces taken for magnetic fields between 0.4 to 5 T, shown in Fig. 5(d): In the curves measured below a magnetic field of between 1 T and 2 T every second quantized plateau is missing, i.e., the filling factor of the plateaus changes by 2. Further, there are clear transitions in Fig. 5(d) between regions in which plateaus with odd multiples of the conductance quantum $e^2/h$ prevail and ones with even multiples.

The experimental observation suggests the picture presented below to explain the checkerboard like pattern in the fan chart and the unique odd-to-even plateau transitions. Figure 6(c) displays a cartoon showing the coexisting electron hole LLs together with the corresponding filling factors. Both electrons and holes are characterized by a partial filling factor $\nu_i$, counting the number of occupied electron and hole LLs. The total filling factor $\nu$ (and the corresponding value of $\sigma_{xy}$ in units of $e^2/h$) is the sum of electron and hole filling factors, $\nu_e$ and $-\nu_h$, respectively. Since the hole LLs are doubly (spin) degenerate, $\nu_h$ can only have even values in contrast to the topological surface electron, for which the filling factor increases in steps of one when $E_F$ is swept across an electron LL. Whether the total filling factor is even or odd therefore depends on the parity of $\nu_e$. Figure 6(d) illustrates this scenario: between the second and the third hole LL, $-\nu_h = -6$, while between the first and the second electron LL, $\nu_e = 1$. Added together, the total filling factor is $\nu = -5$. To understand the experimental result, yet another ingredient is needed. The rate at which electron and hole LLs are filled is greatly different. The values of the partial filling rates $dn_s/dV_g$ and $dp_s/dV_g$ depend both on $B$ and $V_g$, while their sum, the total filling rate, is a constant given by $C/e$. Within a simple picture, the average value of the partial filling rates (averaging over the small oscillatory part) follows the zero magnetic field filling rates. The filling rate at $B = 0$ is much smaller for surface electrons than for holes due to their lower density of states. This is directly shown by the experimental data in Fig. 1(e) [see also Fig. S2(a) and (b)] when $E_F$ is located in the valence band. This also causes a different filling of electron and hole LLs. While for a gate voltage change of, say $\Delta V_g$, a spin-degenerate hole LL with carrier density $2eB/h$ gets fully filled, an electron LL is still largely empty. Or, viewing it the other way round: A change of the electron filling factor $\nu_e$ by 1 is accompanied by changing $\nu_h$ by several even integers (due to the hole LL's spin degeneracy). Within the picture developed above, we can readily explain the experimental observation: A sequence of quantum Hall steps with only odd quantized step occurs, e.g., when the Fermi level stays between the first and second electron LL while it crosses several hole LLs as $V_g$ is varied. When $E_F$ crosses an electron LL the parity of the filling factors changes and an even sequence of quantum Hall steps emerges. The switching between even and odd plateau sequences is direct proof, that the surface states are topological in nature: with conventional spin-degenerate electron states, only Hall conductances with even multiples of $e^2/h$ would appear in experiment. The checkerboard pattern in $\sigma_{xx}$ is a consequence of the joint filling factor. Minima in $\sigma_{xx}$ occur at integer total filling factors, corresponding to the dark blue regions in Fig. 5(b) and 5(e). These minima develop inside the quadrilaterals formed by the electron and hole LLs, see e.g. Fig. 6(c). Figures 6(b) and (d) show model calculations which confirm the arguments presented above and reproduce the peculiar pattern in $\sigma_{xx}$ and $\sigma_{xy}$. Figure 6(d) displays calculated $\sigma_{xy}$ traces for the different charge carrier species in the system. The total Hall conductivity at constant $B = 0.8$ T displays, as in experiment, transitions from odd to even sequences of quantum Hall steps. Further, the calculated fan chart in Fig. 6(b) reproduces qualitatively the checkerboard pattern observed in experiment. We highlight in this fan chart the electron and hole Landau levels. Each of the quadrilaterals contains an integer total filling factor. There is, however, one obvious difference between experiment and calculation: At the crossing points of the LLs in Fig. 6(b) the total density of states is highest, reflected by the bright yellow color. In the experiment shown in Fig. 5(b), in contrast, the value of the conductivity at the crossing points of electron and hole LLs is between maximum and minimum values. The reason is that we ignore anti-crossings in our model. The anti-crossing of LLs reduces the density of states at the crossing points and thus the conductivity. The cartoon in Fig. 6(e) shows the idealized LL crossing assumed in our simplistic model compared to a more realistic one.

At higher magnetic fields, beyond the first electron LL, the checkerboard pattern disappears and the regular fan chart of the total charge carrier density, which emanates from the charge neutrality point, develops. Above about 3T the spin-degeneracy of the bulk holes starts to get resolved. This is shown in more detail by Fig. S3(a) in [30].

**The Conduction band.**

In the conduction band, labeled as region 5 in the Fig. 3(b), the system has common features both with the gap and the VB. Again, bulk and surface carriers coexist in the conductance band. However, the differences in effective masses and filling rates between bulk and surface carriers are much smaller than on the valence band side. In addition, the contribution of bulk carriers to the conductivity is smaller than that of the surface states, which is in contrast to the VB. And, very importantly, the charge sign is the same, resulting in the same direction (i.e. towards positive gate voltages) of the LL fan charts. Nonetheless, due to the coexistence of quantized top surface, bottom surface and bulk electrons an intricate Landau fan chart develops. Clearly visible is the Landau fan chart of the conduction band (bulk) electrons in Fig. 3(b) [see also Fig. S4(d)], which emerges at $E_C$. Further we can identify the Landau levels which originate at the CNP and reflect the total filling factor $\nu$, i.e. the sum of all electrons in the system. These Landau levels dominate at high magnetic field and are the reason why high field SdH oscillation reflect the total carrier density. The corresponding LL fan is highlighted in Fig. S4(b) in [30]. This is in contrast to the Landau fan chart of the top surface electrons, shown in Fig. S4(c), which dominates at lower $B$ and echoes the density of the top surface electrons via the associated SdH oscillations. These Landau levels, extrapolated to their origin, merge at the virtual points of zero density of the top surface. The intermixing of LLs in this regime prevents a clear observation of the Landau fan of the bottom surface in Fig. 3(b). When bulk electrons (doubly degenerate) and Dirac surface electrons (non-degenerate) coexist, we expect, as in the valence band, even-odd transitions of the quantized Hall plateaus. Indeed, we see even-odd switching in Fig. S3(b), but due to the large filling factors the signature is less pronounced as in the case of coexisting electrons and holes.

**Summary.**

In this work, we studied Shubnikov-de Haas oscillations and the quantum Hall effect under the peculiar conditions of a two-dimensional gas of Dirac fermions "wrapped" around the 80 nm thick bulk of a strained HgTe film. The interplay of four different carrier types, i.e. bulk electrons and holes, as well as top and bottom surface Dirac electrons leads to an intricate pattern of the Landau level fan charts. We identified six different regions in the charts, which differ in terms of condition (low or quantizing magnetic field) and types of charge carrier which coexist at the Fermi level. The latter is tuned by means of the applied gate voltage. A simple model based on the superposition of conductivities and densities of states connected to the different subsystems enables us to describe all chart regions qualitatively correct. In particular, a two component Drude model fits the system in weak magnetic fields. The opposite limit of strong magnetic fields is characterized by the QHE state and resolved LLs, where the total charge density defines the position of longitudinal conductivity minima and plateaus in Hall conductivity. In intermediate magnetic fields, the system becomes highly sensitive to the exact position of the Fermi level. In the valence band, we discovered periodic transitions from even to odd filling factors, which we associate with interacting spin-degenerate holes and non-degenerate surface electron LLs. Similar, but less marked behavior we found in the conduction band. In the bulk energy gap, the nature of SdH oscillations is mainly determined by LLs stemming from electrons on the upper surface sitting on a monotonous background conductivity of semi-classical electrons on the bottom surface. The onset of bulk Landau levels observed at $E_c$ in Fig. 3(b) directly shows that the Fermi level, in contrast to previous claims, can be easily tuned into the conduction band [12]. Though we use high-mobility strained HgTe as model system, we note that our conclusions and analyses are valid for a wide class of topological insulators, e.g. Bi-based ones, even if those show, due to their significantly higher disorder, usually much less resolved quantum transport features.

**Acknowledgments.**

The work was supported by the Deutsche Forschungsgemeinschaft (within Priority Programm SPP 1666 "Topological Insulators"), the Elitenetzwerk Bayern Doktorandenkolleg (K-NW-2013-258, "Topological Insulators") and the European Research Council under the European Union's Horizon 2020 research and innovation programme (grant agreement No 787515, "ProMotion"). D.K. acknowledges support from the Russian Scientific Foundation (Project No. 18-72-00189).

**Appendix. Computer Simulation of Landau Levels Filling**

In order to explain the LL fan chart, obtained from $\sigma_{xx}$ and $\sigma_{xy}$ data, we developed a model which qualitatively describes the system consisting of several groups of carriers in quantizing magnetic fields and at zero temperature. Within the model, the conductances of each group *i* of carriers are independent and add up to the total $\sigma_{xx}$ and $\sigma_{xy}$. All charge carrier species share the same Fermi level $E_F$, determined by the total charge $Q$, which depends linearly on the gate voltage $V_g$ but not on the magnetic field. The output of the calculation are both, partial and total filling factors, the densities of states (DOS) $D^i$ and Hall conductivities $\sigma_{xy}{}^i$ for particular $B$ and $V_g$ values. To achieve semi-quantitative agreement with experiment we need to take the system's electrostatics into account.

The following simplifications were used in the model. Depending on the gate voltage, up to four groups of charge carriers are present in the system, namely top and bottom Dirac surface electrons, bulk electrons (conduction band) and bulk holes (valence band), marked by the indices *top*, *bot*, *b* (bulk) and *h* (holes) below. Each of these groups is characterized by a parabolic two-dimensional dispersion law with an effective mass of $m_i$ (values used in the calculation are listed in Table. 1). This approach neglects the quasi-linear Dirac dispersion of the surface states, but is justified as we seek qualitative understanding rather than full quantitative agreement. Next, each group is characterized by its lowest energy $E_c^i$ (Dirac point for top and bottom electrons, conduction band edge for bulk electrons) and valence band edge $E_v$ (for holes). The electron (hole) density in each group reads $n_s^i = \int_{E_c^i}^{E_F} D^i(E)dE$ ($p_s = \int_{E_F}^{E_v} D^h(E)dE$). While at zero magnetic field the constant DOS is given by $D^i = g_s^i m_i/(2\pi\hbar^2)$ ($g_s^i$= spin degeneracy, $\hbar$=reduced Planck constant) at non-zero $B$ Landau levels emerge, described by a sum of Gaussians with linewidth $\Gamma^i$:

$$D^i(E) = \frac{g_{LL}}{\Gamma^i\sqrt{\pi}} \sum_{n=0}^{\infty} e^{-\frac{(E-E_n^i)^2}{\Gamma^2}}.$$

Here $g_{LL} = eB/h$ is the spin-resolved LL degeneracy and $E_n$ the energy of the $n^{th}$ LL. For surface carriers, LLs are spin-resolved and $E_n^i = E_c^i + \hbar\omega_c^i\left(n + \frac{1}{2}\right)$, with $\omega_c^i = eB/m_i$ the cyclotron frequency holds. For bulk carriers each LL is initially doubly degenerate, but the degeneracy is lifted due to Zeeman splitting, $\Delta_Z^i = g_i^*\mu_B B$, where $g_i^*$ is the Landé g-factor of bulk carriers (see table 1). Thus, for bulk electrons, $E_n^\pm = E_c^b + \hbar\omega_c^b\left(n + \frac{1}{2}\right) \pm \frac{\Delta_Z^b}{2}$ and for holes, $E_n^\pm = E_v - \hbar\omega_c^h\left(n + \frac{1}{2}\right) \pm \frac{\Delta_Z^h}{2}$, holds.

The electrostatics of the system is introduced in the same way as done in our previous work [23], but extended to the case of quantizing magnetic fields. Each group of carriers is treated as an infinitely large, charged capacitor plate of zero thickness with charge density $\Sigma = -en_s^i$ ($ep_s$ for holes), located at certain distances from the gate. Every charged plate induces a symmetric electric field $F = \Sigma^i/2\varepsilon\varepsilon_0$ on both sides of the plate, which, in turn, creates an electrostatic energy shift for the other groups of carriers. On the other hand, the electric field below the bottom surface is zero. Therefore, the charge on the bottom surface creates an electric field of $en_s^{bot}/\varepsilon_{HgTe}\varepsilon_0$ inside the HgTe layer, where $\varepsilon_0$ is the electric vacuum permittivity and $\varepsilon_{HgTe} = 80$, the effective (see below and "the limitation of the model" section) dielectric constant of HgTe. Then, the electrostatic energy shift for a layer located in a distance $d$ from the bottom is $e^2 n_s^{bot} d/\varepsilon_{HgTe}\varepsilon_0$. In our model, a change of the electrostatic energy shifts the position of $E_c^i$ and $E_v$. The energy shift for bulk carriers is solely determined by the charge on the bottom surface (because there are no other carriers in between), while the top surface electrons are affected by both bottom surface electrons and bulk carriers with weights proportional to the respective distance. The dielectric constant of the HgTe layer was chosen in a way to get agreement with the partial filling factors of top and bottom surface electrons, found in experiment. To the same end, the (average) spatial location of bulk electrons and holes was taken as a fitting parameter (see Table 1). In order to agree not only with filling rates, but also with the experimentally extracted carrier densities, an additional constant energy shift was introduced for each group except the one on the bottom surface.

The calculation procedure was as follows: First, for a specified $V_g$ value the total charge in the system is calculated by the linear relation $Q = -\alpha (V_g - V_g^{CNP})$, where $\alpha = C/e = 2.42 \cdot 10^{11}$ cm$^2$/V·s is the total filling rate. Next, for a given value of the magnetic field the partial densities of states $D^i(E)$ are calculated analytically taking the electrostatic energy shifts into account. Finally, the equation $Q = ep_s(E_F, E_v) - e\sum_i n_s^i(E_F, E_c^i)$ is solved numerically giving the values of $E_F$, $n_s^i$ and $p_s$ as an output. With these values the partial densities of states, $\sigma_{xx}$ and $\sigma_{xy}$ were calculated and plotted in Fig. 6.

Table. 1. Parameters of the different groups of carriers.

| carrier group | top surface electrons | bottom surface electrons | bulk electrons | bulk holes |
|---|---|---|---|---|
| density notation | $n_s^{top}$ | $n_s^{bot}$ | $n_s^b$ | $p_s$ |
| density value at CNP | 4.1·10$^{10}$ cm$^{-2}$ | 2.9·10$^{10}$ cm$^{-2}$ | 0 | 7·10$^{10}$ cm$^{-2}$ |
| charge | -e | -e | -e | e |
| $m_i$ | $0.03 \cdot m_0$ | $0.03 \cdot m_0$ | $0.06 \cdot m_0$ | $0.3 \cdot m_0$ |
| $g_s^i$ | 1 | 1 | 2 | 2 |
| $g_s^*$ | - | - | 33.3 | 6.7 |
| $\Delta_Z / \hbar\omega_c$ | - | - | 0.3 | 0.3 |
| $\Gamma\|_{B=1T}$ | 1 meV | 1.5 meV | 0.5 meV | 0.1 meV |
| $d$ (distance from bottom surface) | 80 nm | - | 52 nm | 75 nm |

**The limitation of the model.**

As stated before, the model was developed in order to achieve a qualitative agreement with the experimental data, but, surprisingly even semi-quantitative concordance is achieved in some regions. In particular, the DOS oscillations in the bulk energy gap fit the experimentally obtained conductivity oscillations well. Excellent agreement is also achieved in the limit of strong magnetic fields. The main benefit of the model is that it explains conclusively – while utmost simple – the observed magnetotransport peculiarities. However, the model has a set of built-in limitations, which are described below:

1. The real band dispersion at zero magnetic field is replaced for all kinds of carriers by a parabolic one. This reduction is justified, because the QHE is of fundamental nature and only slightly depends on the details of the band structure. The observed magnetotransport features are explained by the interplay of the LLs of different charge carriers, taking the system's electrostatics and LL broadening due to disorder into account. Using the real band structure will simply lead to a renormalization of the broadening parameters but does not lead to any new physics. We note that the use of a parabolic dispersion significantly simplifies the model.

2. The effect of the out of plane electric field on the electronic band dispersion, expected in thin HgTe films [19], is neglected. One of the main consequences of the band distortion is a reduced partial filling rate for the bottom side surface electrons (and an increased one for top surface electrons, accordingly). In order to fit the experimentally obtained filling rates, we have therefore to use an effective dielectric constant of the HgTe layer of $\varepsilon_{HgTe} = 80$, which is larger than the real one of about 20.

3. No quantum mechanical interaction between LLs is taking into account. When two LLs (stemming from bulk holes and surface electrons, e.g.) cross, a maximum in the DOS forms in our calculations. In a more realistic scenario, the interaction between carriers should lead to an anti-crossing of LLs and reduced values of the DOS and the longitudinal conductivity at the crossing. As a consequence we cannot reproduce all details of the checkerboard pattern shown, e.g., in Fig. 5(b).

4. In the limit of strong magnetic fields, the experimentally measured width of the conductivity peaks is smaller than the width of gap region on the gate voltage scale. In contrast, our DOS calculations show minima and maxima regions of similar width. This is a well-known discrepancy, which is due to the fact that (i) the conductivity in the quantum Hall regime is proportional to the square of the Landau level DOS [2] and that (ii) effects which stem from transport in edge states (parallel to the applied field) are not taken into account.

5. The valley degeneracy of bulk holes is neglected. Band structure calculations predict a 4-fold degeneracy for (100)-oriented HgTe films and QWs and a 2-fold one for (013) oriented ones, used here. However, to our knowledge the valley degeneracy has not been observed in experiment yet. Some authors believe that the valley degeneracy is lifted because of the Rashba effect [37].

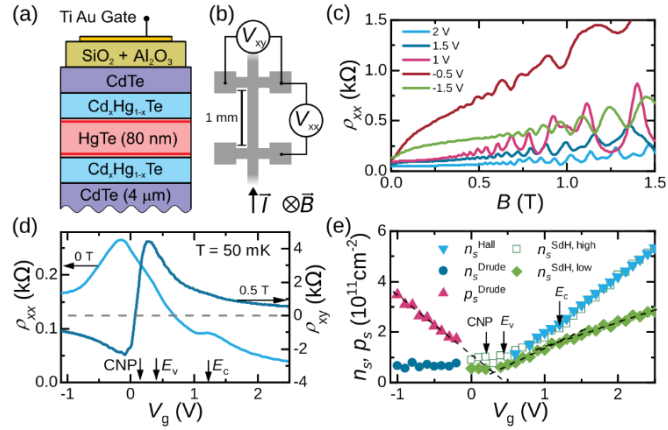

**Figure 1**: (a) Scheme of the heterostructure cross section. (b) Sketch of the 200 μm wide and 1100 μm long Hall bar. (c) Longitudinal resistivity $\rho_{xx}(B)$ at different $V_g$ for $E_F$ in the conduction band (2, 1.5 V), gap (1 V), and valence band (-0.5, -1.5 V). (d) $V_g$-dependence of $\rho_{xx}$ (left axis, light blue) at $B = 0$ and $\rho_{xy}$ (right axis, dark blue) at $B = 0.5$ T. The vertical arrows mark the charge neutrality point (CNP), top of the bulk valence ($E_V$) and bottom of the conductance band ($E_C$), respectively. (e) Electron $n_s(V_g)$ and hole densities $p_s(V_g)$ extracted from the Hall slope, two-carrier Drude model and SdH oscillations. The analysis follows Ref. [13] and yields positions of the valence and conduction band edges at $V_g(E_V) = 0.45$ V and $V_g(E_C) = 1.2$ V, respectively, marked in (d).

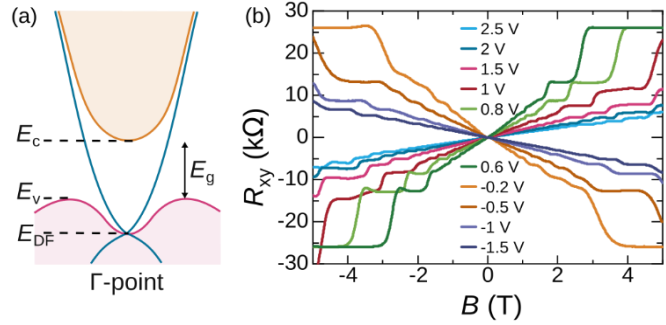

**Figure 2**: (a) Schematic band structure of strained HgTe. Conduction and valence band edges are marked by $E_C$ and $E_V$, the Dirac point, which is located in the valence band, by $E_{DF}$. The topological Dirac surface states are shown in blue. Note, that the position of $E_F$ within the band structure is usually different for top and bottom surface. (b) The quantum Hall effect in strained 80 nm HgTe, shown for different gate voltages $V_g$, can stem from surface electrons in the gap (with additional contributions from bulk electrons in the CB) and bulk holes in the VB. This is reflected by the changing sign of the Hall slope when tuning the gate voltage $V_g$ across the charge neutrality point.

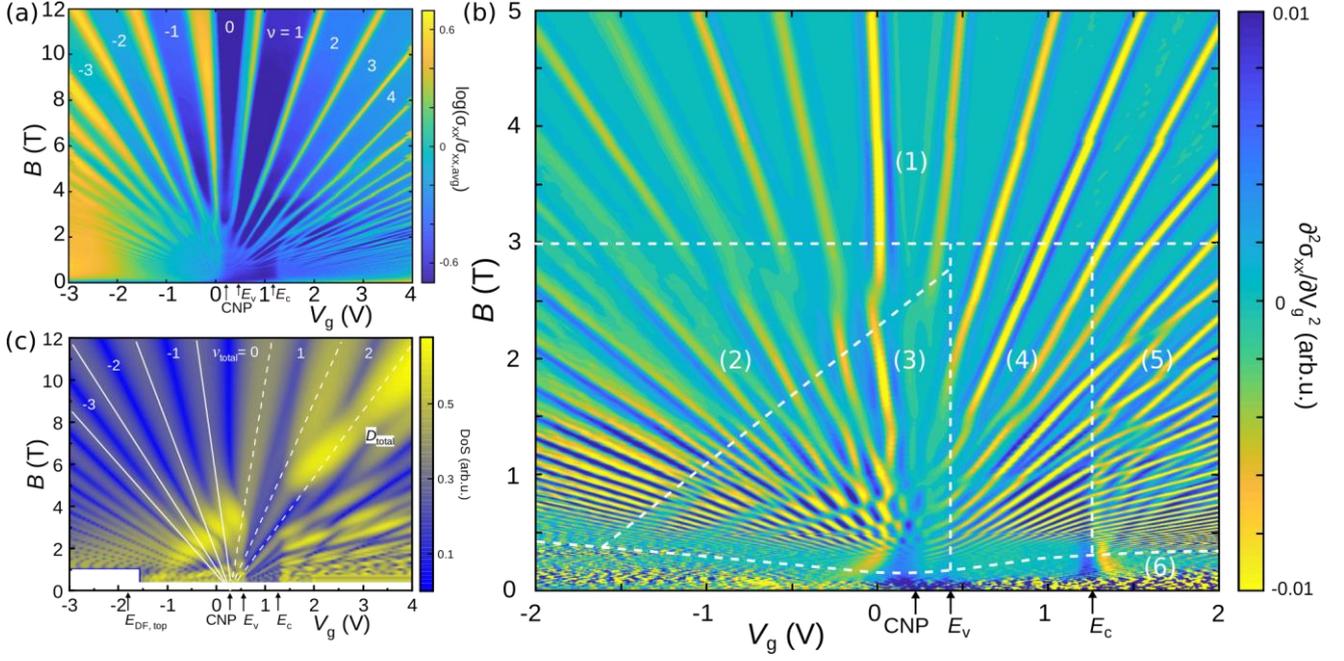

**Figure 3**: (a) Color map of the normalized longitudinal conductivity $\sigma_{xx}(V_g, B)$ for magnetic fields up to $B = 12$ T. Conductance maxima are yellow/green; filling factors for the blueish energy gaps are shown in white. Data is normalized to the average value for a given $B$ value. We use a logarithmic color scale to enhance the visibility of features throughout the entire investigated parameter space. (b) Excerpt of the data in (a), shown as the 2$^{nd}$ derivative $\partial^2 \sigma_{xx}/\partial V_g^2$. Yellow lines correspond, as above, to Landau levels and green/blue regions to Landau gaps. White dashed lines separate the investigated parameter space $(V_g, B)$ into distinct sections discussed in the text. (c) Simulated density of states based on the simple model described in the Appendix section. Maxima (minima) of the DoS are shown in yellow (blue) and correspond to maxima (minima) of $\sigma_{xx}(V_g, B)$, i.e. to LL positions (gaps). $E_V$, $E_C$, CNP and $E_{DF,top}$ label points on the gate voltage axis, which were matched to the experimentally obtained values. White lines represent the fan chart originating from the CNP.

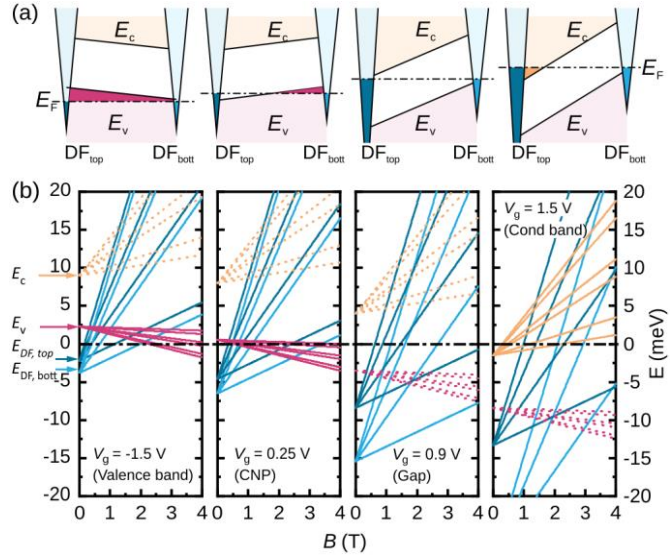

**Figure 4**: When a gate voltage is applied, the charge carriers of the system rearrange. While the exact solution requires a complicated self-consistent solution of Schrödinger and Poisson equations, most of the features observed in magnetotransport can be explained by a simple electrostatic model. Within this model, each group of carriers is modeled by a 2D system of electrons or holes of zero thickness. Each layer generates its own set of Landau levels emerging at different energies, but with common Fermi level position. Charged 2D surfaces cause, like a charged capacitor plate, 1D electric fields across the structure leading to electrostatic potentials, depending linearly on the charge. These potentials change with gate voltage, thus shifting the band edges, from which the Landau levels emanate. (a) Schematic band diagrams of the investigated HgTe 3D TI system at different gate voltages $V_g$ (VB: $V_g = -1.5$ V; CNP: $V_g = 0.25$ V; Gap: $V_g = 0.9$ V; CB: $V_g = 1.5$ V). Occupied hole states are shown in red, occupied states of surface electrons in dark blue and bulk states in orange. The Dirac points of top and bottom surfaces in strained HgTe thin films are located below $E_V$, as shown in the diagrams. (b) Landau level spectra at different bias conditions used in the model. The corresponding band bending is shown above in (a). The model calculations include Zeeman-splitting for spin-degenerate bulk states (orange for conduction band, red for valence band states). Landau fans that do not cross the Fermi energy $E_F$ (i.e. empty levels) are shown as dotted lines.

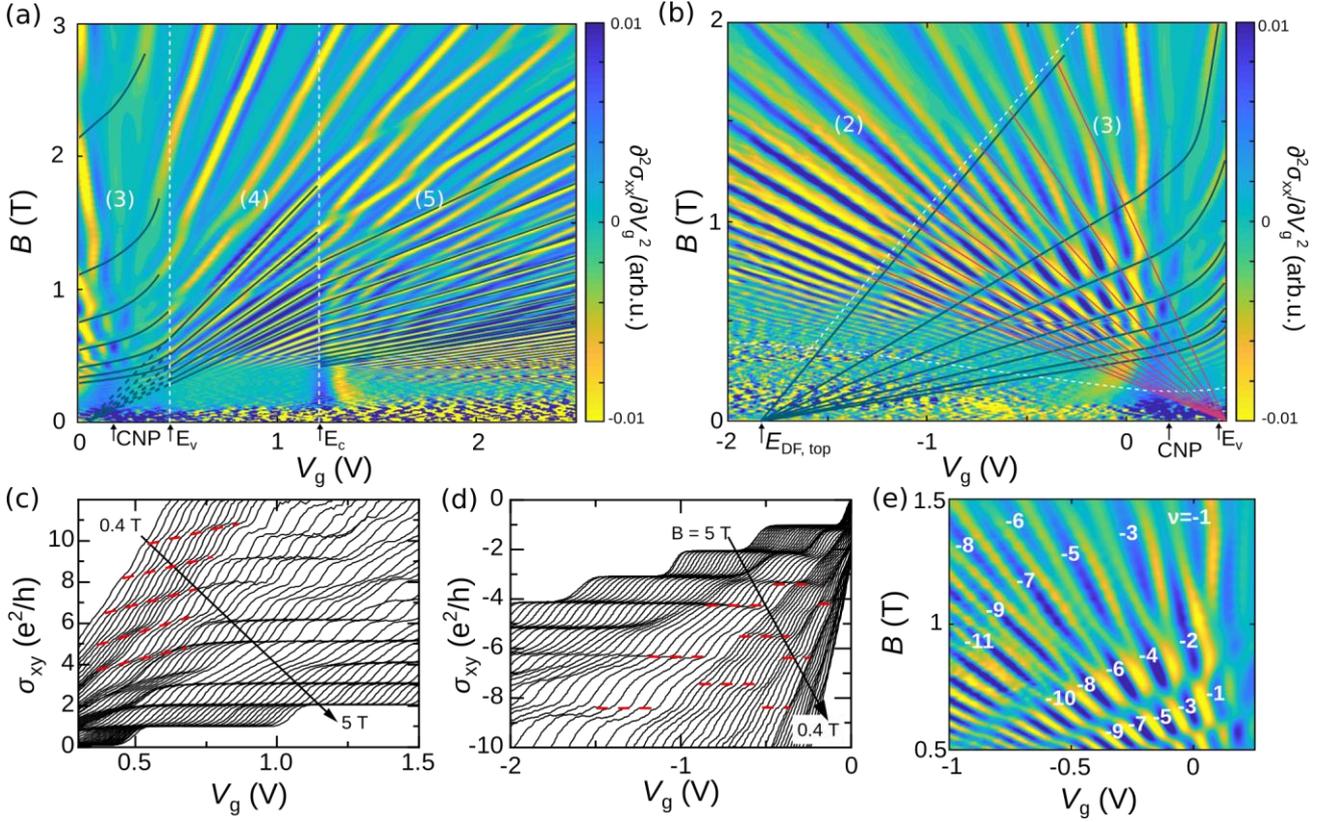

**Figure 5**: (a) $\partial^2\sigma_{xx}/\partial V_g^2$ for $B$ up to 3 T and $V_g > 0$ covering the bulk gap and conduction band region. Dark blue lines retrace the Landau levels. The Landau fan in the gap between $V_g(E_V)$ and $V_g(E_C)$ has its virtual origin at $V_g = 0.05$ V and stems from the top surface electrons. A distinct change of slope is observed when $E_F$ enters the bulk bands at $V_g(E_V)$ and $V_g(E_C)$. This is due to the reduced filling rates of surface electrons in the bulk bands. The distortion of the Landau fan in region 4 at fields between 1 T and 2 T is ascribed to the onset of quantization of the bottom surface electrons. (b) $\partial^2\sigma_{xx}/\partial V_g^2$ for $E_F$ in the valence band, i.e., for $V_g < V_g(E_V)$ and $B$ up to 2 T. Landau fans stemming from bulk holes (red) and top surface electrons (black) coexist in this regime. Whenever a bulk hole LL crosses a surface electron LL, the total filling factor parity changes, resulting in a shift of the SdH phase. (c) Hall conductivity $\sigma_{xy}(V_g)$ for $B = 0.4..5$ T and $E_F$ in the bulk gap. For small $B$-fields the plateau values of the individual traces increase with increasing field (dashed lines) due to superposition of quantized Hall conductivity from the top surface (constant $\sigma_{xy}$) and classical Hall conductivity of back surface electrons ($\sigma_{xy}$ linear in $V_g$). At higher $B$ quantized steps appear. (d) The Hall conductivity $\sigma_{xy}(V_g)$ measured in the valence band. In the region of the checkerboard pattern alternating sequences of plateaus with only odd/even filling factors (dashed lines) show up. The changes in filling factor parity stem from coexisting spin-resolved topological surface states and spin-degenerate bulk holes. (e) Enlarged region of coexisting electron and hole Landau levels of (b) showing the alternating sequences of even (odd) total filling factors.

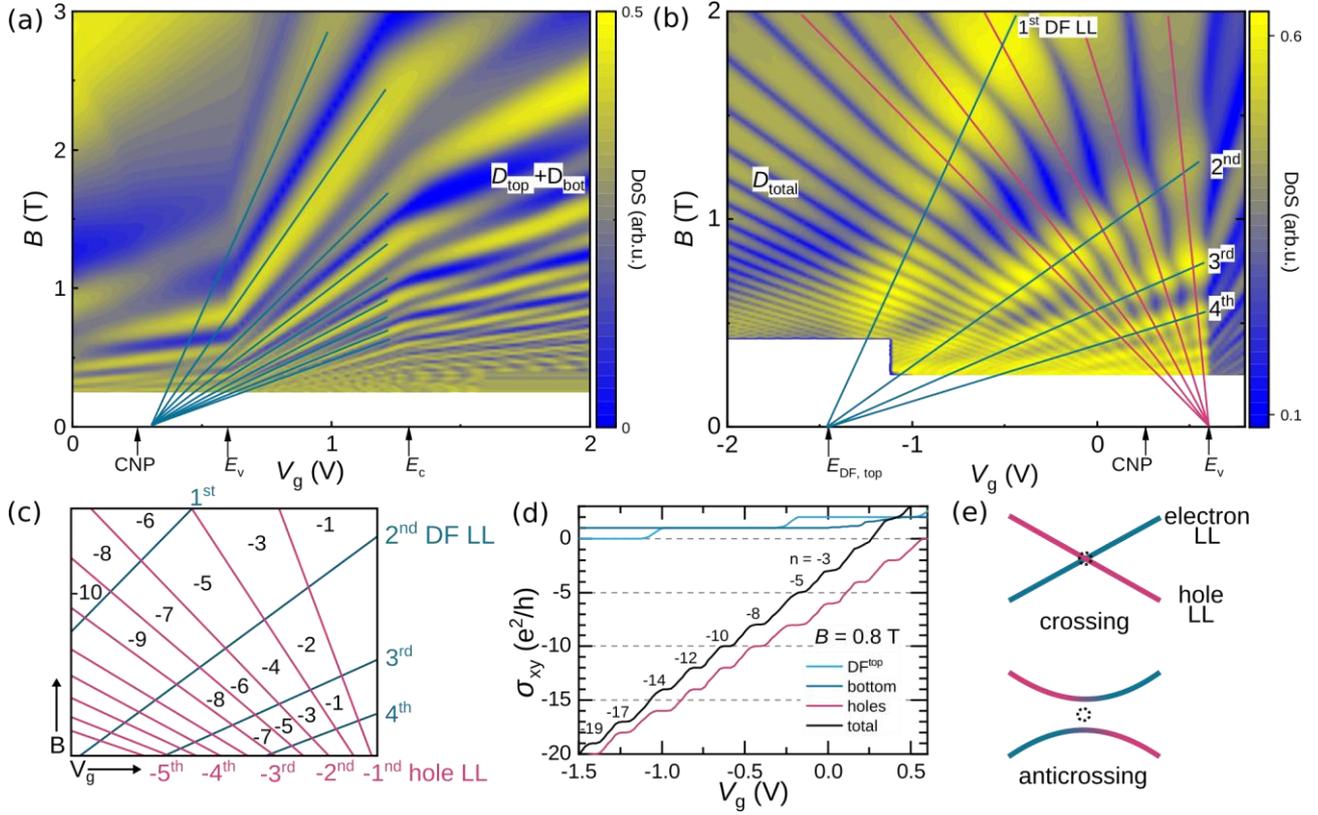

**Figure 6**: (a) Calculated partial DOS of top and bottom surface states (extracted from the full calculation described in the Appendix) near the bulk gap and for $B \leq 3$ T. Yellow stands for high DOS values (i.e. LLs) while the blue color represents small DOS (i.e. the gaps between LLs). $E_V$, $E_C$, CNP and $E_{DF,top}$ labels points on the gate voltage axis, which were matched to experimentally obtained values. The observed LLs exhibit, as in experiment, pronounced kinks at the band edges ($E_V$, $E_C$) resulting from the abrupt change of the partial filling rates when bulk states get filled. (b) The calculated total DOS in the valence band, corresponding to the experiment shown in Fig. 5(c) shows, for low $B$, an unusual pattern stemming from the interplay of spin-degenerate hole LLs (red lines) and non-degenerate electron LLs (blue lines). Color code as in (a). (c) Sketch of the crossing electron (Dirac fermion) (blue) and hole LLs (red). When $E_F$ crosses a Dirac fermion LL, the filling factor changes by 1, when it crosses a hole LL by 2. The black numbers give the resulting filling factors in between electron and hole LLs. (d) $\sigma_{xy}$ as a function of $V_g$ calculated for the different charge carrier fractions when degenerate and non-degenerate hole and electron LLs, as in (b), coexist. The total Hall conductivity $\sigma_{xy}$ (black line) shows, as in experiment, the switching between odd and even plateau values, depending on the parity of the electron filling factor. (e) The simulations shown here only considers the electrostatics of the LLs resulting in degenerate LLs at crossing points. However, quantum mechanics requires that such crossings be avoided. This situation is sketched here to show that in our model LL crossings lead (incorrectly) to maxima in the DOS while in experiment such crossings are characterized by a reduced density of states at the crossing point (black circle).

[1]  K. V. Klitzing, G. Dorda, and M. Pepper, Phys. Rev. Lett. **45**, 494 (1980).

[2]  T. Ando, A. B. Fowler, and F. Stern, Rev. Mod. Phys. **54**, 437 (1982).

[3]  R. Dingle, H. L. Störmer, A. C. Gossard, and W. Wiegmann, Appl. Phys. Lett. **33**, 665 (1978).

[4]  K. S. Novoselov, E. McCann, S. V. Morozov, V. I. Fal'ko, M. I. Katsnelson, U. Zeitler, D. Jiang, F. Schedin, and A. K. Geim, Nat. Phys. **2**, 177 (2006).

[5]  E. E. Mendez, L. Esaki, and L. L. Chang, Phys. Rev. Lett. **55**, 2216 (1985).

[6]  K. Suzuki, K. Takashina, S. Miyashita, and Y. Hirayama, Phys. Rev. Lett. **93**, 016803 (2004).

[7]  M. Karalic, C. Mittag, S. Mueller, T. Tschirky, W. Wegscheider, K. Ensslin, T. Ihn, and L. Glazman, Phys. Rev. B **99**, 201402 (2019).

[8]  M. Z. Hasan and C. L. Kane, Rev. Mod. Phys. **82**, 3045 (2010).

[9]  Y. Ando, J. Phys. Soc. Japan **82**, 102001 (2013).

[10] J. E. Moore, Nature **464**, 194 (2010).

[11] X. L. Qi and S. C. Zhang, Rev. Mod. Phys. **83**, 1057 (2011).

[12] C. Brüne, C. X. Liu, E. G. Novik, E. M. Hankiewicz, H. Buhmann, Y. L. Chen, X. L. Qi, Z. X. Shen, S. C. Zhang, and L. W. Molenkamp, Phys. Rev. Lett. **106**, 126803 (2011).

[13] D. A. Kozlov, Z. D. Kvon, E. B. Olshanetsky, N. N. Mikhailov, S. A. Dvoretsky, and D. Weiss, Phys. Rev. Lett. **112**, 196801 (2014).

[14] Y. Xu, I. Miotkowski, C. Liu, J. Tian, H. Nam, N. Alidoust, J. Hu, C. K. Shih, M. Z. Hasan, and Y. P. Chen, Nat. Phys. **10**, 956 (2014).

[15] R. Yoshimi, A. Tsukazaki, Y. Kozuka, J. Falson, K. S. Takahashi, J. G. Checkelsky, N. Nagaosa, M. Kawasaki, and Y. Tokura, Nat. Commun. **6**, 6627 (2015).

[16] W. Zou, W. Wang, X. Kou, M. Lang, Y. Fan, E. S. Choi, A. V. Fedorov, K. Wang, L. He, Y. Xu, and K. L. Wang, Appl. Phys. Lett. **110**, 212401 (2017).

[17] C. Thomas, O. Crauste, B. Haas, P. H. Jouneau, C. Bäuerle, L. P. Lévy, E. Orignac, D. Carpentier, P. Ballet, and T. Meunier, Phys. Rev. B **96**, 245420 (2017).

[18] L. Fu and C. L. Kane, Phys. Rev. B **76**, 045302 (2007).

[19] S. C. Wu, B. Yan, and C. Felser, Europhys. Lett. **107**, 57006 (2014).

[20] C. Brüne, C. Thienel, M. Stuiber, J. Böttcher, H. Buhmann, E. G. Novik, C. X. Liu, E. M. Hankiewicz, and L. W. Molenkamp, Phys. Rev. X **4**, 041045 (2014).

[21] J. Zhu, C. Lei, and A. H. MacDonald, ArXiv E-Prints 1804.01662 (2018).

[22] K. M. Dantscher, D. A. Kozlov, P. Olbrich, C. Zoth, P. Faltermeier, M. Lindner, G. V. Budkin, S. A. Tarasenko, V. V. Bel'kov, Z. D. Kvon, N. N. Mikhailov, S. A. Dvoretsky, D. Weiss, B. Jenichen, and S. D. Ganichev, Phys. Rev. B **92**, 165314 (2015).

[23] D. A. Kozlov, D. Bauer, J. Ziegler, R. Fischer, M. L. Savchenko, Z. D. Kvon, N. N. Mikhailov, S. A. Dvoretsky, and D. Weiss, Phys. Rev. Lett. **116**, 166802 (2016).

[24] H. Maier, J. Ziegler, R. Fischer, D. Kozlov, Z. D. Kvon, N. Mikhailov, S. A. Dvoretsky, and D. Weiss, Nat. Commun. **8**, 2023 (2017).

[25] J. Ziegler, R. Kozlovsky, C. Gorini, M. H. Liu, S. Weishäupl, H. Maier, R. Fischer, D. A. Kozlov, Z. D. Kvon, N. Mikhailov, S. A. Dvoretsky, K. Richter, and D. Weiss, Phys. Rev. B **97**, 035157 (2018).

[26] D. A. Kozlov, J. Ziegler, N. N. Mikhailov, S. A. Dvoretskii, and D. Weiss, JETP Lett. **109**, 799 (2019).

[27] G. M. Gusev, D. A. Kozlov, A. D. Levin, Z. D. Kvon, N. N. Mikhailov, and S. A. Dvoretsky, Phys. Rev. B **96**, 045304 (2017).

[28] J. G. Checkelsky, L. Li, and N. P. Ong, Phys. Rev. Lett. **100**, 206801 (2008).

[29] A. J. M. Giesbers, L. A. Ponomarenko, K. S. Novoselov, A. K. Geim, M. I. Katsnelson, J. C. Maan, and U. Zeitler, Phys. Rev. B **80**, 201403 (2009).

[30] See Supplemental Material at http://link.aps.org/supplemental/..., (2020).

[31] Z. D. Kvon, E. B. Olshanetsky, D. A. Kozlov, N. N. Mikhailov, and S. A. Dvoretskii, JETP Lett. **87**, 502 (2008).

[32] A. F. Young, C. R. Dean, L. Wang, H. Ren, P. Cadden-Zimansky, K. Watanabe, T. Taniguchi, J. Hone, K. L. Shepard, and P. Kim, Nat. Phys. **8**, 550 (2012).

[33] G. M. Minkov, V. Y. Aleshkin, O. E. Rut, A. A. Sherstobitov, A. V. Germanenko, S. A. Dvoretski, and N. N. Mikhailov, Phys. Rev. B **96**, 035310 (2017).

[34] X. C. Zhang, I. Martin, and H. W. Jiang, Phys. Rev. B **74**, 073301 (2006).

[35] G. Yu, D. J. Lockwood, A. J. Springthorpe, and D. G. Austing, Phys. Rev. B **76**, 085331 (2007).

[36] C. A. Duarte, G. M. Gusev, A. A. Quivy, T. E. Lamas, A. K. Bakarov, and J. C. Portal, Phys. Rev. B **76**, 075346 (2007).

[37] K. Ortner, X. C. Zhang, A. Pfeuffer-Jeschke, C. R. Becker, G. Landwehr, and L. W. Molenkamp, Phys. Rev. B **66**, 075322 (2002).

# Supplementary Information for

# "Quantum Hall effect and Landau levels in the 3D topological insulator HgTe"

J. Ziegler[1], D.A. Kozlov[2,3], N.N. Mikhailov[3], S. Dvoretsky[3], D. Weiss[1]

[1] Experimental and Applied Physics, University of Regensburg, D-93040 Regensburg, Germany
[2] A.V. Rzhanov Institute of Semiconductor Physics, Novosibirsk 630090, Russia
[3] Novosibirsk State University, Novosibirsk 630090, Russia

## S1. Supplemental figures

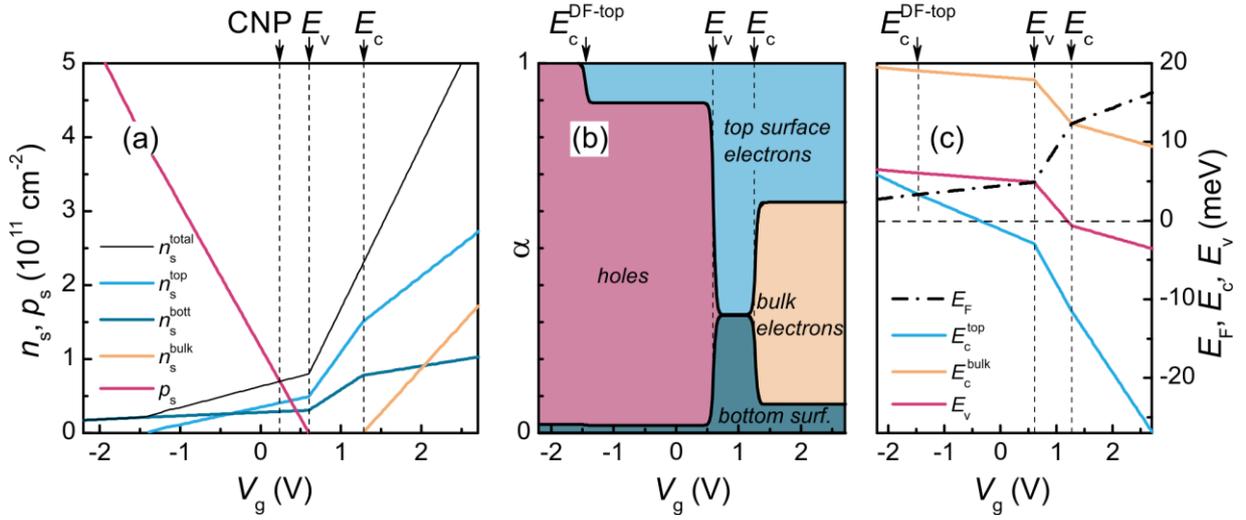

**Figure S1**: Supplemental figures on the electrostatics. (a) The charge carrier densities of the total system ($n_s^{total}$), the top ($n_s^{top}$) and bottom surfaces ($n_s^{bot}$), bulk electron ($n_s^{bulk}$) and hole states ($p_s$) obtained from the simulations to our data as a function of $V_g$. The filling rates $\alpha$ (given by the slopes of the $V_g$-dependencies) of surface carriers change when crossing $V_g(E_C)$ or $V_g(E_V)$. (b) The ratio of filling factors $\alpha$ of the different carrier species as a function of $V_g$. (c) The shifts of the Fermi energy $E_F$ and the band edges $E_C^{top}$, $E_C^{bulk}$, and $E_V$ due to the electrostatics are visualized.

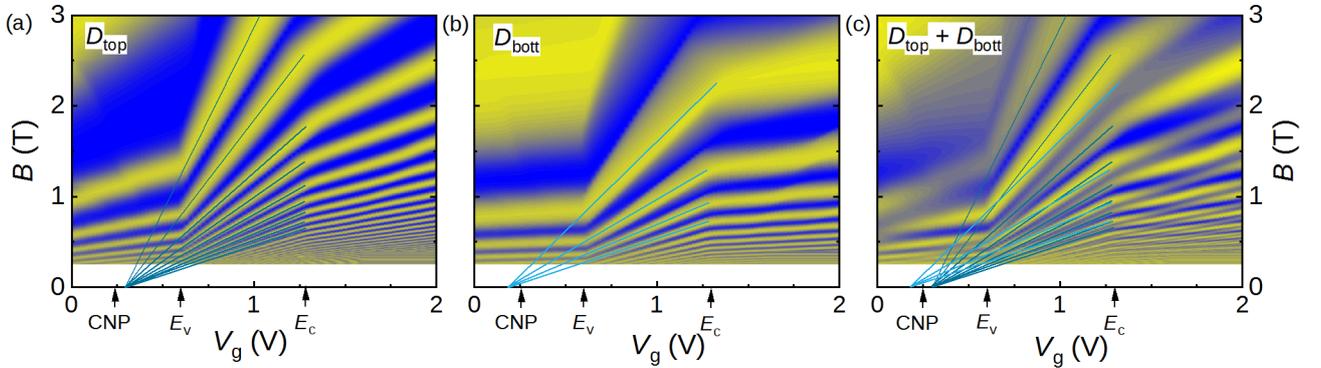

**Figure S2**: Partial density of states DOS near the bulk gap region extracted from the full simulation for (a) the top surface, (b) the bottom surface, and (c) their sum. While both surfaces show distinct LL formation, fitted by respective the Landau fans, the top surface Landau fan dominates the combined DOS in (c). The LLs exhibit pronounced kinks at the borders of the energy gap ($E_V$ and $E_C$) due to abrupt changes of the partial filling rates in the presence of bulk carriers.

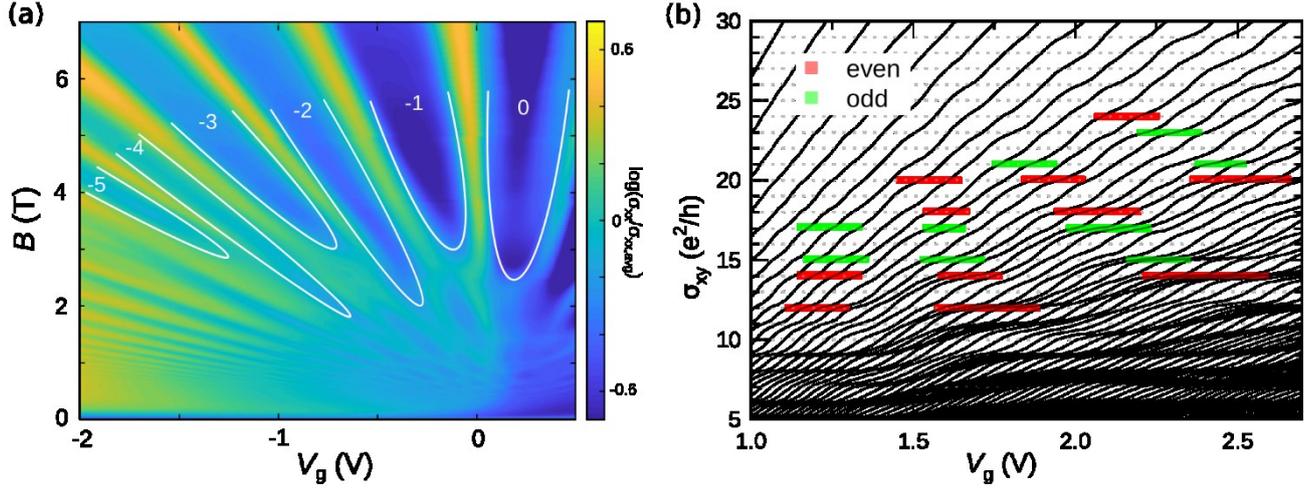

**Figure S3**: (a) Zoom-in of data from Fig. 3(a), showing the normalized conductivity $\sigma_{xx}/\sigma_{xx,avg}$ up to 7 T in the valence band. At fields below 3 T, the alternating even and odd SdH sequences can be seen. Filling factors $\nu$ at higher fields are labeled in white. SdH minima with even filling factors are found to persist down to smaller fields than odd filling factors. This indicates that the Landau gap exceeds the Zeeman gap. (b) Hall conductivity traces $\sigma_{xy}(V_g)$ in the conduction band demonstrate similar features as found in the valence band [shown in Fig. 4(d)]. Distinct even and odd plateau sequences are less pronounced due to the much higher total filling factors.

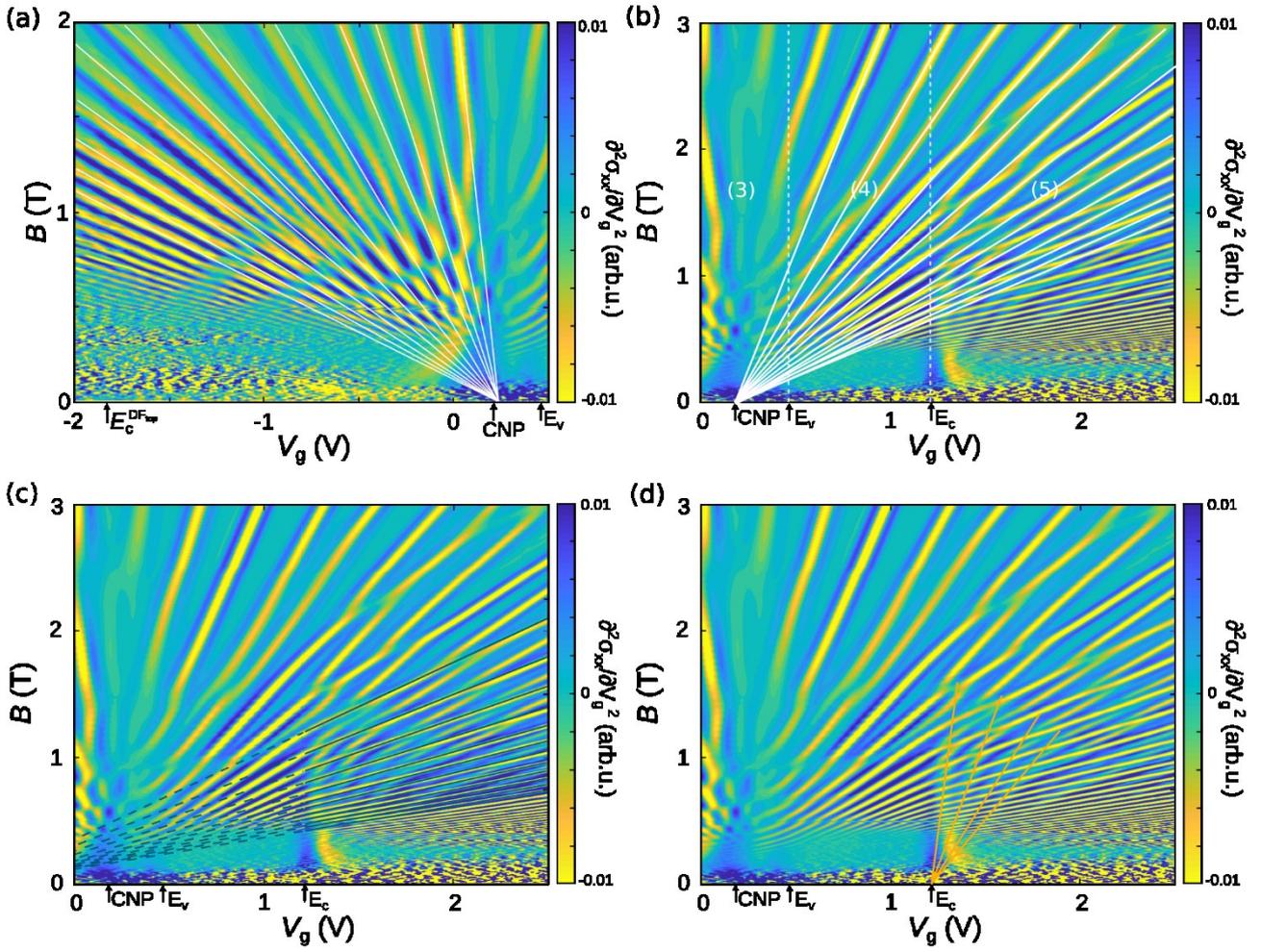

**Figure S4**: Color plots of the same data we show in Fig. 4 (a) and (b), with the different Landau fans mentioned in the main text. (a) Landau fan for holes originating at the CNP (i.e. regions 2 and 3) belong to the total filling factor $n_s - p_s$ (white lines). (b) Landau fan for electrons originating at the CNP, shown by white lines for the total electron density (bulk plus surface density). (c) Landau fan (dark lines) for the top surface electrons only, which emanates at the virtual Dirac point (not shown in the plot). The dashed lines extrapolate the fan into the gap and the valence band region where different filling rates prevail. (d) Landau fan of the conduction band electrons (orange lines) which start at the conduction band edge $E_C$.

## S2. Estimation of the diffusive bottom surface

Here we estimate the contribution of the diffusive bottom surface to the conductivity in the energy gap (region 4) at small magnetic fields between 0.7 and 1 T, shown in Fig. 5(b). The top surface shows quantized behavior with well-defined plateaus in the conductivity, so that $\sigma_{xy}^{top} = \nu e^2/h$. Transport on the bottom surface is still diffusive at these fields, so that it can be described with the Drude model:

$$\sigma_{xy}^{bot} = \sigma_0 \frac{\mu B}{1+\mu^2 B^2} \text{ with } \sigma_0 = e n_s \mu.$$

Here, $\mu$ is the charge carrier mobility, $B$ the magnetic field and $n_s$ the charge carrier density. The total conductivity of the system is given by the sum, $\sigma_{xy} = \sigma_{xy}^{top} + \sigma_{xy}^{bot}$. While the contribution of the top surface is constant, the bottom surface depends on $n_s$. Exploiting the system's electrostatics, $\Delta n_s/\Delta V_g = \alpha^{bot}$, where $\alpha^{bot}$ is the filling rate of the bottom surface, we get

$$\frac{\Delta \sigma_{xy}^{bot}}{\Delta V_g} = \alpha^{bot} \frac{e \mu^2 B}{1+\mu^2 B^2}.$$

From characterization measurements shown in Fig. 1(e), we extract $\alpha^{bot}$ = 0.93·10$^{11}$ cm$^{-2}$V$^{-1}$ and the charge carrier mobility. Typical values for the mobilites, when transport is surface dominated in the bulk gap, are $\mu$ = 400 000 cm$^2$/Vs. The calculation yields values of $\Delta \sigma_{xy}/\Delta V_g$(1 T) = 3.8 e$^2$/h V$^{-1}$ and $\Delta \sigma_{xy}/\Delta V_g$(0.7 T) = 5.5 e$^2$/h V$^{-1}$. Slopes extracted from $\sigma_{xy}$ data in Fig. 5(b) range from 3.1 e$^2$/h V$^{-1}$ and 4.1 e$^2$/h V$^{-1}$, thus showing reasonable agreement.